\RequirePackage{lineno}
\documentclass[reprint,showpacs,amsmath,amssymb,aps,prl,floatfix,preprintnumbers]{revtex4-1}
\usepackage{graphicx}
\usepackage{dcolumn}
\usepackage{bm}
\usepackage{hyperref}
\usepackage{placeins}

\newcommand{\infb}{fb$^{-1}$}
\newcommand{\lumi}{$4.9$\;\infb}
\newcommand\T{\rule{0pt}{2.6ex}}
\newcommand\B{\rule[-1.2ex]{0pt}{0pt}}
\newcommand{\mycomment}[1]{}

\newcommand{\myfontsizetwo}{\fontsize{8}{10}\selectfont}
\def\ptt{\ensuremath{p_{\mathrm{Tt}}}}

\begin{document}

\preprint{CERN-PH-EP-2012-013}
\preprint{Submitted to Physical Review Letters}
\vspace*{0.2cm}
\title{Search for the Standard Model Higgs boson in the diphoton decay\\channel with \lumi\ of pp collision data at $\sqrt{s}=$7~TeV with ATLAS}

\collaboration{The ATLAS Collaboration}

\begin{abstract}
A search for the Standard Model Higgs boson is performed in the diphoton decay channel. The data used corresponds to an integrated luminosity of \lumi\, collected with the ATLAS detector at the Large Hadron Collider in proton-proton collisions at a center-of-mass energy of $\sqrt{s}=7\,\mathrm{TeV}$. In the diphoton mass range $110{\--}150$~GeV, the largest excess with respect to the background-only hypothesis is observed at 126.5~GeV, with a local significance of 2.8 standard deviations. Taking the look-elsewhere effect into account in the range $110{\--}150$~GeV, this significance becomes 1.5 standard deviations.
The Standard Model Higgs boson is excluded at 95\% confidence level in the mass ranges of $113{\--}115$~GeV and $134.5{\--}136$~GeV.
\end{abstract}

\maketitle

\hyphenation{ATLAS}

The Higgs mechanism~\cite{HiggsAll} is one of the best-motivated processes to explain electroweak (EW) symmetry breaking. In the Standard Model (SM), this mechanism explains the generation of the $W$ and $Z$ boson masses and predicts the existence of the only elementary scalar in the SM, the hypothetical Higgs boson. Prior direct searches at LEP, Tevatron and LHC exclude the SM Higgs boson with a mass $m_H <114.4$~GeV and $145<m_H<206$~GeV at 95\% confidence level~(CL)~\cite{lep,Tevatron,Atlaslnln}. The present search for $H\to\gamma\gamma$ uses the full 2011 data sample collected by ATLAS at 7~TeV center-of-mass energy and updates prior results with 1.08 $\rm{fb}^{-1}$~\cite{ATLASHtoGamGam2011}.

The ATLAS detector~\cite{ATLAS_detector} consists of an inner tracking detector surrounded by a superconducting solenoid providing a 2\,T magnetic field, electromagnetic and hadron calorimeters, and a muon spectrometer. The main subdetectors relevant to the search presented here are the calorimeters, in particular the electromagnetic section, and the inner tracking system. The inner detector provides tracking in the pseudorapidity region $|\eta| < 2.5$
and consists of silicon pixel- and microstrip-detectors inside a transition radiation tracker. The electromagnetic calorimeter, a lead/liquid-argon sampling device, is divided in one barrel ($|\eta|<1.475$) and two end-cap ($1.375<|\eta|<3.2$) sections.
The barrel ($|\eta|<0.8$) and extended barrel ($0.8<|\eta|<1.7$) hadron calorimeter sections consist
of steel and scintillating tiles, while the end-cap sections ($1.5<|\eta|<3.2$) are composed of copper and liquid argon.

The data were recorded using a diphoton trigger~\cite{Triggerpaper}, each photon having a transverse energy, $E_{\rm{T}}$, of at least 20~GeV, seeded by a lower-level trigger that required two clusters in the electromagnetic calorimeter with $E_{\rm{T}}>$12 or 14~GeV, depending on the data-taking period. The trigger efficiency for the signal events passing the final offline selection is 99\%.
After applying data quality requirements, the total integrated luminosity of the dataset used in this analysis is $4.9\pm0.2\;\rm{fb}^{-1}$~\cite{LumiAll}.

Events are required to contain at least one vertex with at least three associated tracks, where the transverse momentum, $p_{\rm{T}}$, of each track is required to be larger than 0.4~GeV, as well as two photon candidates each seeded by an energy cluster in the electromagnetic calorimeter with $E_{\rm{T}}>2.5$~GeV.
Photons that convert to electron-positron pairs in the inner detector leave one or two tracks that are reconstructed and matched to the clusters in the calorimeter. The photon energy is calibrated separately for converted and unconverted photon candidates using Monte Carlo (MC) simulations of the detector~\cite{CSCNote}. A correction, depending on pseudorapidity and typically of the order of $\pm 1\%$, is applied to the calibrated photon energy as obtained from studies using $Z\to ee$ decays in data~\cite{ElecPerfPaper2010}. Photons are reconstructed in the fiducial region $|\eta|<2.37$, excluding the calorimeter barrel/end-cap transition regions $1.37<|\eta|<1.52$. The photon candidates are ordered in $E_{\rm{T}}$ and the leading (subleading) candidate is required to have $E_{\rm{T}}>40$~GeV (25~GeV).
Both candidates are required to pass further identification criteria based on shower shapes measured in the electromagnetic calorimeter and on the energy leakage into the hadron calorimeter~\cite{InclusiveCrossSection}. The photon reconstruction and identification efficiency ranges typically from 65\% to 95\% for $E_{\rm{T}}$ in the range 25 to 80~GeV. The two photon candidates are required to be isolated by having at most 5~GeV energy deposited in the calorimeters in a cone of $\Delta R=\sqrt{(\Delta \eta)^2 + (\Delta \phi)^2}=0.4$ around the candidate, where $\phi$ is the azimuthal angle, after subtracting the energy assigned to the photon itself. The measured isolation~\cite{InclusiveCrossSection} is corrected for lateral shower leakage and ambient energy from multiple proton-proton interactions (pileup), following the method in Ref.~\cite{Cacciari:area}.
The isolation cut retains $\sim87\%$ of Higgs boson signal events with $m_H=120$~GeV while rejecting $\sim44\%$ of the selected data, which includes jets that can be misidentified as photons.

The opening angle of the two photons, used in the calculation of their invariant mass, is determined using the trajectories of the photons. For a converted photon with a well-measured conversion vertex, the trajectory is determined from the straight line between the barycenter of the associated energy deposits in the calorimeter and the conversion vertex. Otherwise, the trajectory is determined from the barycenters of the showers in the first and second layers of the calorimeter. The extrapolation of the trajectories as well as the average beam spot position are used to determine the origin of the photons along the beam axis, $z$. The resolution of the $z$ vertex coordinate is $\sim6$\,mm on average for two converted photons with reconstructed tracks, and $\sim15\,$mm otherwise.
The contribution of the resulting angular resolution to the mass resolution is negligible in comparison to that of the energy resolution.

In total 22489 events pass the selection in the diphoton mass range $100{\--}160$~GeV. To confirm the dominance of the diphoton processes ($\gamma\gamma$) over backgrounds with one or two misidentified jets ($\gamma$j, jj), the composition of the selected sample is estimated using the data. A sideband technique~\cite{ATLASHtoGamGam2011} is used to estimate the numbers of $\gamma\gamma$, $\gamma$j or jj events. The fraction of true diphoton events is estimated to be $\left(71\pm5\right)\%$. The amount of Drell-Yan background is estimated by selecting $Z\to ee$ decays in data where either one or both electrons pass the photon selection. The measured composition is summarized in Table~\ref{tab:composition} and is compatible with MC expectations.
This decomposition is not directly used in the signal search; however, it is used to validate the parametrization of the background fit (see below).
\begin{table}[h!]
\centering
{\myfontsizetwo
\caption{Composition of the selected sample as obtained from the data in the mass window of $100{\--}160$~GeV. A sum in quadrature of statistical and systematic uncertainties is quoted.}
\label{tab:composition}
\begin{tabular}[c]{lcccc}
\hline\hline
\T \B          & $\gamma\gamma$            & $\gamma j$             & $jj$                     & Drell-Yan \\
\hline
\T \B Events   & $16000\pm1100$            & $5230\pm890$            & $1130\pm600$       & $165\pm8$ \\
\T \B Fraction & $\left(71\pm5\right)\%$   & $\left(23\pm4\right)\%$ & $\left(5\pm3\right)\%$  & $\left(0.7\pm0.1\right)\%$  \\
\hline\hline
\end{tabular}
}
\end{table}

The events are separated into nine mutually exclusive categories with different mass resolutions and signal-to-background ratios, to increase the sensitivity to a possible Higgs boson signal. Categories are defined by the conversion status, $\eta$ of the selected photons, and \ptt~\footnote{$p_{\rm{Tt}} = {|\vec{p}_{\rm{T}}^{\gamma\gamma} \times \widehat{t}|}$, where ${\widehat{t} = \frac{ {\vec{p}_\mathrm{T}^{\gamma_1}} -{\vec{p}_\mathrm{T}^{\gamma_2}} }{ |{\vec{p}_\mathrm{T}^{\gamma_1}} - {\vec{p}_\mathrm{T}^{\gamma_2}}|}}$ denotes the transverse thrust, ${\vec{p}_\mathrm{T}^{\gamma_1}}$ and ${\vec{p}_\mathrm{T}^{\gamma_2}}$ are the transverse momenta of the two photons, and ${\vec{p}_{\rm{T}}^{\gamma\gamma}={\vec{p}_\mathrm{T}^{\gamma_1}} + {\vec{p}_\mathrm{T}^{\gamma_2}}}$ is the transverse momentum of the diphoton system.}, the component of the diphoton pT that is orthogonal to the thrust axis, as proposed in Ref.~\cite{pTtpaper}. Events with two unconverted photons are separated into \emph{unconverted central} ($|\eta|<0.75$ for both candidates) and \emph{unconverted rest} (all other events). Events with at least one converted photon are separated into \emph{converted central} ($|\eta|<0.75$ for both candidates), \emph{converted transition} (at least one photon with $1.3<|\eta|<1.75$) and \emph{converted rest} (all other events). Excepting the \emph{converted transition} category, each category is further divided by a cut at \ptt=40~GeV into two categories, \emph{low} \ptt\ and \emph{high} \ptt. MC studies show that signal events, particularly those produced in vector-boson fusion (VBF) or in associated production ($W/ZH$ and $t\bar{t}H$), have on average larger \ptt\ than background events. The number of data events in each category is given in Table~\ref{tab:SBnumbers}.
\begin{table}[t!]
\caption{Mass resolution~$\sigma_\mathrm{CB}$ (see text) and FWHM (both in GeV), expected number of signal events~($\rm{N_S}$) for $m_H=120$~GeV and number of events in the data~($\rm{N_D}$) in each category for \lumi. $\rm{N_S}$ and $\rm{N_D}$ are for the mass range $100{\--}160$~GeV. The signal-to-background ratios (S/B) are given in a mass window containing 90\% of the signal for $m_H=120$~GeV.}
\label{tab:SBnumbers}
\myfontsizetwo
\begin{center}
\begin{tabular}{lccccr}
\hline\hline
\T \B Category & $\sigma_\mathrm{CB}$ & FWHM & $\rm{N_S}$ & $\rm{N_D}$ & S/B \\
\hline
\T \B Unconverted central, low \ptt   &  1.4  & 3.4 & 9.1 & 1763   & 0.05 \\
\T \B Unconverted central, high \ptt  &  1.4  & 3.3 & 2.6 & 235    & 0.11 \\
\T \B Unconverted rest, low \ptt      &  1.7  & 4.0 & 17.7 & 6234  & 0.02 \\
\T \B Unconverted rest, high \ptt     &  1.6  & 3.9 & 4.7 & 1006   & 0.04 \\
\T \B Converted central, low \ptt     &  1.6  & 3.9 & 6.0 & 1318   & 0.03 \\
\T \B Converted central, high \ptt    &  1.5  & 3.6 & 1.7 & 184    & 0.08 \\
\T \B Converted rest, low \ptt        &  2.0  & 4.7 & 17.0 & 7311  & 0.01 \\
\T \B Converted rest, high \ptt       &  1.9  & 4.5 & 4.8 & 1072   & 0.03 \\
\T \B Converted transition            &  2.3  & 5.9 & 8.5 & 3366   & 0.01 \\
\hline
\T \B All categories                  &  1.7  & 4.1 & 72.1 & 22489 & 0.02 \\
\hline\hline
\end{tabular}
\end{center}
\end{table}

\begin{figure}[h!]
\includegraphics[width=.49\textwidth]{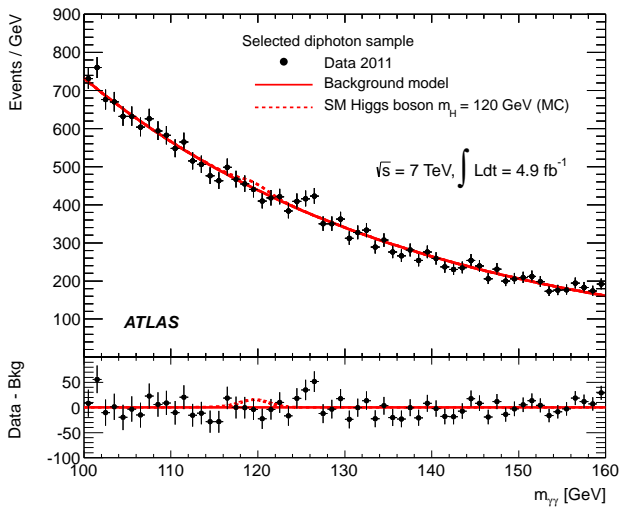}
\caption{\label{fig:inclusiveData} Invariant mass distribution for the selected data sample, overlaid with the total background (see text). The bottom inset displays the residual of the data with respect to the total background. The Higgs boson expectation for a mass hypothesis of 120~GeV corresponding to the SM cross section is also shown.}
\end{figure}

The distribution of the invariant mass of the diphoton events, $m_{\gamma\gamma}$, summed over all categories, is shown in Fig.~\ref{fig:inclusiveData}. The sum of the background-only fits (described below) to the invariant mass in each of the categories is superimposed. The signal expectation for a SM Higgs boson with $m_H=120$~GeV is also shown. The presence of the Higgs boson will appear as a narrow resonance in the invariant mass of the selected photon pairs superimposed on a smoothly falling background. The residual of the data with respect to the total background as a function of $m_{\gamma\gamma}$ is also shown in Fig.~\ref{fig:inclusiveData}.

Higgs boson production and decay are simulated with several MC samples that are passed through a full detector simulation~\cite{atlasSim} using {\sc Geant4}~\cite{geant4}. Pileup effects are simulated by overlaying each MC event with a variable number of MC inelastic proton-proton collisions~\cite{MC11Tune}.
{\sc Powheg}~\cite{POWHEGALL}, interfaced to {\sc Pythia}~\cite{PYTHIA} for showering and hadronization, is used for generation of gluon fusion and VBF production. {\sc Pythia} is used to generate the Higgs boson production in association with $W/Z$ and $t\bar{t}$.

The Higgs boson production cross sections are computed up to next-to-next-to-leading order (NNLO)~\cite{a1a2GGF} in QCD for the gluon fusion process. In addition, QCD soft-gluon resummations up to next-to-next-to-leading log (NNLL) improve the NNLO calculation~\cite{Catani:2003zt}. The next-to-leading order (NLO) EW corrections are applied~\cite{a3GGFNLOEWALL}. These results are compiled in Refs.~\cite{a4GGFPROGRAMSALL} assuming factorisation between QCD and EW corrections.
The cross sections for the VBF process are calculated with full NLO QCD and EW corrections~\cite{VBFNLO}, and approximate NNLO QCD corrections are available~\cite{Bolzoni:2010xr}.
The $W/ZH$ processes are calculated at NLO~\cite{Han:1991ia} and at NNLO~\cite{Brein:2003wg}, and NLO EW radiative corrections~\cite{Ciccolini:2003jy} are applied.
The full NLO QCD corrections for $t\bar{t}H$ are calculated~\cite{TTHALLNEW}.
The Higgs boson cross sections, branching ratios~\cite{BRALL} and their uncertainties are compiled in Ref.~\cite{LHCXSECALL}.

The cross sections multiplied by the branching ratio into two photons are listed in Table~\ref{tab:signal}. The number of signal events produced by gluon fusion is rescaled to take into account the expected destructive interference between the $gg\to\gamma\gamma$\; continuum background and the $gg\to H\to\gamma\gamma$\; process~\cite{Dixon}, leading to a reduction of the production rate by $2{\--}5\%$ depending on $m_H$ and analysis category. The fractions of gluon-fusion, VBF, $WH$, $ZH$ and $t\bar{t}H$ production are approximately $87\%$, $7\%$, $3\%$, $2\%$ and $1\%$, respectively, for $m_H=120$~GeV.

The shower shape variables of the simulated samples are shifted to agree with the corresponding distributions in the data~\cite{InclusiveCrossSection} and the photon energy resolution is broadened to account for differences observed between $Z\to ee$ data and MC events.
Events generated with {\sc{Powheg}} at NLO have been reweighted to match the Higgs boson $p_{\mathrm{T}}$ distribution predicted by {\sc HqT}~\cite{deFlorian:2011xf}. The signal yields expected for \lumi\ and selection efficiencies are given in Table~\ref{tab:signal}.

\begin{figure}[h!]
\begin{center}
\includegraphics[width=0.45\textwidth]{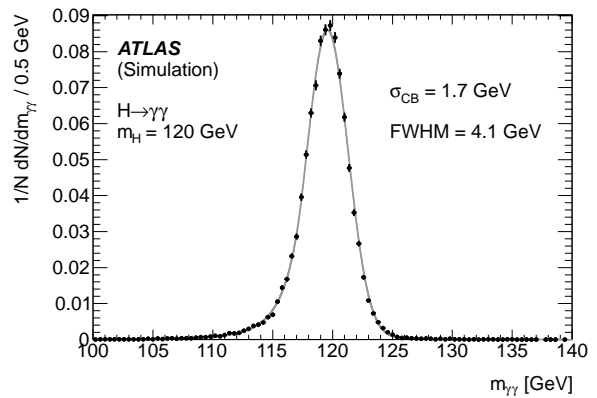}
\end{center}
\caption{Reconstructed invariant mass distribution for a simulated signal of $m_H=120$~GeV summed over all categories, superimposed with the fit to the signal model.}
\label{fig:signalfit}
\end{figure}

The invariant mass shape of the signal in each category is modeled by the sum of a Crystal Ball function~\cite{CSB} describing the core of the distribution with a width $\sigma_{CB}$, and a wide Gaussian with a small amplitude describing the tails of the mass distribution. In Fig.~\ref{fig:signalfit} the sum of all signal processes in all categories is shown for a Higgs boson with $m_H=120$~GeV. The expected full-width-at-half-maximum (FWHM) is 4.1~GeV and $\sigma_{CB}$ is 1.7~GeV. The resolution varies with category (see Table~\ref{tab:SBnumbers}). The signal-to-background ratio (S/B), calculated in a mass window symmetric about the signal maximum and containing 90\% of the signal, varies from 0.11 to 0.01 depending on the category and is also shown in Table~\ref{tab:SBnumbers}.

The background in each category is estimated from the data by fitting the diphoton mass spectrum in the range $100{\--}160$~GeV with an exponential function with free slope and normalization parameters. The background curve in Fig.~\ref{fig:inclusiveData} is the sum of these nine contributions. For each category, a single exponential fit satisfactorily describes the mass spectrum. This has been checked using large samples of diphoton events produced by the {\sc Resbos}~\cite{Resbos} and {\sc Diphox}~\cite{diphox} MC generators.

The difference between the exponential function and the true background will contribute
to an excess or a deficit of events over background expectations.  In order to take this
into account in a conservative way, a term is included in the likelihood function that allows
for a signal-like component that is consistent with the background uncertainty. For each category this uncertainty is estimated from MC by the difference between the mass distribution of diphoton events generated with {\sc Resbos} and the result of the exponential fit to this distribution. Photon reconstruction and identification efficiencies are taken into account. The MC events are scaled to correspond to $4.9~\rm{fb}^{-1}$ of data. The uncertainty is then the maximal difference between the MC shape and the model integrated in a sliding mass window of 4~GeV, the approximate FWHM of the expected signal. The uncertainties obtained are $\pm(0.1{\--}7.9)$ events depending on the category. Pseudo experiments are used to check that the sum of $\gamma\gamma$, $\gamma$j and jj events can also be described well by the exponential model.
The background uncertainties are further validated by fitting the data with functions that have more degrees of freedom than the single exponential, and comparing the residuals to those obtained with the exponential fit.

\begin{table}[t!]
\caption{Higgs boson production cross section multiplied by the branching ratio into two photons, expected number of signal events summed over all categories for \lumi\ and selection efficiencies for various Higgs boson masses.}
\label{tab:signal}
\begin{center}
{\myfontsizetwo
\begin{tabular}{lccccccccc}
\hline\hline
\T \B $m_H$ [GeV] & 110 & 115 & 120 & 125 & 130 & 135 & 140 & 145 & 150  \\
\hline
\T \B $\sigma\times BR$ [fb]  & 45 & 44 & 43 & 40 & 36 & 32 & 27 & 22 & 16 \\
\T \B Signal events           & 69 & 72 & 72 & 69 & 65 & 58 & 50 & 41 & 31 \\
\T \B Efficiency [\%]         & 31 & 33 & 34 & 35 & 37 & 37 & 38 & 38 & 39 \\
\hline\hline
\end{tabular}}
\end{center}
\end{table}

The dominant experimental uncertainty on the signal yield is the photon reconstruction and identification efficiency ($\pm11\%$), which is estimated with data by using electrons from $Z$ and $W$ decays and photons selected from $Z\to\ell\ell\gamma$ ($\ell=e,\mu$) events. Pileup also affects the identification efficiency and contributes to the uncertainty ($\pm4\%$). Further uncertainties on the signal yield are related to the trigger ($\pm1\%$), Higgs boson $p_{\rm{T}}$ modeling ($\pm1\%$), isolation ($\pm5\%$) and luminosity ($\pm3.9\%$). Uncertainties on the predicted cross sections are due to uncertainties on the QCD renormalization and factorization scales (${}^{+12}_{-8}\%$) and on the parton density functions (PDF, \cite{Botje:2011sn} and references therein) and $\alpha_s$ ($\pm8\%$). The total uncertainty on the signal yield is ${}^{+20}_{-17}\%$. The total uncertainty on the mass resolution is $\pm14\%$, dominated by the uncertainty on the energy resolution of the calorimeter, determined from $Z\to ee$ events ($\pm12\%$). Further uncertainties on the mass resolution result from an imperfect knowledge of material in front of the calorimeter affecting the extrapolation from electron to photon calibration ($\pm6\%$), the impact of pileup ($\pm3\%$) estimated from events taken with random triggers, and the photon angle measurement ($\pm1\%$) estimated using $Z\to ee$ events. The uncertainty on the knowledge of the material in front of the calorimeter is used to derive the amount of event migration between the converted and unconverted categories ($\pm4.5\%$). Different PDFs and scale variations in {\sc HqT} calculations are used to derive possible event migration between high and low \ptt\ categories ($\pm8\%$).

A modified frequentist approach~($CL_S$)~\cite{cls} for setting limits and a frequentist approach to calculate the $p_0$ value are used~\cite{HCG}. The $p_0$ is the probability that the background fluctuates to the observed number of events or higher. The combined likelihood, which is a function of the ratio of the measured cross-section relative to that of the SM prediction, is constructed from the unbinned likelihood functions of the nine categories. Systematic uncertainties are incorporated by introducing nuisance parameters with constraints. Asymptotic formulae~\cite{asym} are used to derive the limits and $p_0$ values, which are refined with pseudo experiments~\cite{combinationLetter}, as functions of the hypothetical Higgs boson mass.

\begin{figure}[htb]
\includegraphics[width=.49\textwidth]{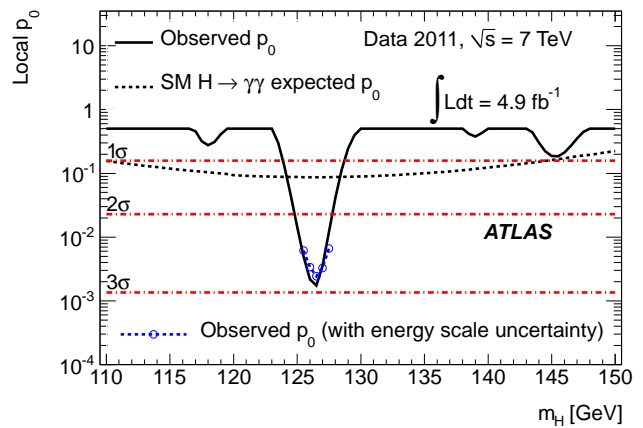}
\caption{The observed local $p_0$, the probability that the background fluctuates to the observed
number of events or higher (solid line). The open points indicate
the observed local $p_0$ value when energy scale uncertainties are taken into
account. The dotted line shows the expected median local $p_0$ for the signal hypothesis
when tested at $m_H$.}
\label{fig:p0}
\end{figure}

\begin{figure}[htb!]
\centering
\includegraphics[width=0.49\textwidth]{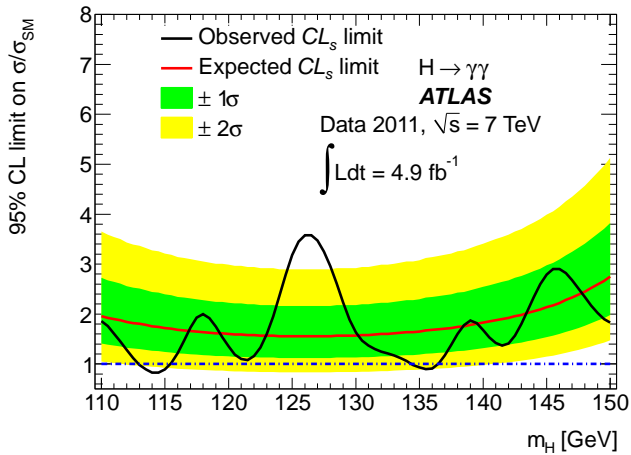}
\caption{Observed and expected 95\% CL limits on the SM Higgs boson production
normalized to the predicted cross section as a function of $m_H$.}
\label{fig:CLS}
\end{figure}

The observed and expected local $p_0$ values and the 95\%~CL limits on the Higgs boson production in units of the SM cross section are displayed in Figs.~\ref{fig:p0} and~\ref{fig:CLS}. Before considering the uncertainty on the signal mass position, the largest excess with respect to the background-only hypothesis in the mass range $110{\--}150$~GeV is observed at 126.5 GeV with a local significance of 2.9 standard deviations. The uncertainty on the mass position ($\pm0.7$~GeV) due to the imperfect knowledge of the photon energy scale has a small effect on the significance. When this uncertainty is taken into account, the significance is 2.8 standard deviations; this becomes 1.5 standard deviations when the look elsewhere effect~\cite{LEE} for the mass range $110{\--}150$~GeV is included. The median expected upper limits of the cross section in the absence of a true signal, at the 95\% CL, vary between 1.6 and 1.7 times the SM cross section in the mass range $115{\--}130$~GeV, and between 1.6 and 2.7 in the mass range $110{\--}150$~GeV. The observed 95\% CL upper limit of the cross section relative to the SM cross section is between 0.83 and 3.6 over the full mass range. A SM Higgs boson is excluded at 95\% CL in the mass ranges of $113{\--}115$~GeV and $134.5{\--}136$~GeV. These results are combined with SM Higgs searches in other decay channels in Ref.~\cite{combinationLetter}.

We thank CERN for the very successful operation of the LHC, as well as the support staff from our institutions without whom ATLAS could not be operated efficiently.

We acknowledge the support of ANPCyT, Argentina; YerPhI, Armenia; ARC, Australia; BMWF, Austria; ANAS, Azerbaijan; SSTC, Belarus; CNPq and FAPESP, Brazil; NSERC, NRC and CFI, Canada; CERN; CONICYT, Chile; CAS, MOST and NSFC, China; COLCIENCIAS, Colombia; MSMT CR, MPO CR and VSC CR, Czech Republic; DNRF, DNSRC and Lundbeck Foundation, Denmark; ARTEMIS and ERC, European Union; IN2P3-CNRS, CEA-DSM/IRFU, France; GNAS, Georgia; BMBF, DFG, HGF, MPG and AvH Foundation, Germany; GSRT, Greece; ISF, MINERVA, GIF, DIP and Benoziyo Center, Israel; INFN, Italy; MEXT and JSPS, Japan; CNRST, Morocco; FOM and NWO, Netherlands; RCN, Norway; MNiSW, Poland; GRICES and FCT, Portugal; MERYS (MECTS), Romania; MES of Russia and ROSATOM, Russian Federation; JINR; MSTD, Serbia; MSSR, Slovakia; ARRS and MVZT, Slovenia; DST/NRF, South Africa; MICINN, Spain; SRC and Wallenberg Foundation, Sweden; SER, SNSF and Cantons of Bern and Geneva, Switzerland; NSC, Taiwan; TAEK, Turkey; STFC, the Royal Society and Leverhulme Trust, United Kingdom; DOE and NSF, United States of America.

The crucial computing support from all WLCG partners is acknowledged gratefully, in particular from CERN and the ATLAS Tier-1 facilities at TRIUMF (Canada), NDGF (Denmark, Norway, Sweden), CC-IN2P3 (France), KIT/GridKA (Germany), INFN-CNAF (Italy), NL-T1 (Netherlands), PIC (Spain), ASGC (Taiwan), RAL (UK) and BNL (USA) and in the Tier-2 facilities worldwide.

\bibliography{paper}
\onecolumngrid

\clearpage
\onecolumngrid
\begin{flushleft}
{\Large The ATLAS Collaboration}

\bigskip

G.~Aad$^{\rm 48}$,
B.~Abbott$^{\rm 110}$,
J.~Abdallah$^{\rm 11}$,
S.~Abdel~Khalek$^{\rm 114}$,
A.A.~Abdelalim$^{\rm 49}$,
A.~Abdesselam$^{\rm 117}$,
O.~Abdinov$^{\rm 10}$,
B.~Abi$^{\rm 111}$,
M.~Abolins$^{\rm 87}$,
O.S.~AbouZeid$^{\rm 157}$
% ...
\bigskip

% ...
$^{10}$ Institute of Physics, Azerbaijan Academy of Sciences, Baku, Azerbaijan\\
$^{11}$ Institut de F\'isica d'Altes Energies and Departament de F\'isica de la Universitat Aut\`onoma  de Barcelona and ICREA, Barcelona, Spain\\
% ...
$^{48}$ Fakult\"{a}t f\"{u}r Mathematik und Physik, Albert-Ludwigs-Universit\"{a}t, Freiburg i.Br., Germany\\
$^{49}$ Section de Physique, Universit\'e de Gen\`eve, Geneva, Switzerland\\
% ...
$^{87}$ Department of Physics and Astronomy, Michigan State University, East Lansing MI, United States of America\\
% ...
$^{110}$ Homer L. Dodge Department of Physics and Astronomy, University of Oklahoma, Norman OK, United States of America\\
$^{111}$ Department of Physics, Oklahoma State University, Stillwater OK, United States of America\\
% ...
$^{114}$ LAL, Univ. Paris-Sud and CNRS/IN2P3, Orsay, France\\
% ...
$^{117}$ Department of Physics, Oxford University, Oxford, United Kingdom\\
% ...
$^{157}$ Department of Physics, University of Toronto, Toronto ON, Canada\\
% ...
\end{flushleft}


\end{document}